\documentclass[a4paper,11pt]{article}
\pdfoutput=1 % if your are submitting a pdflatex (i.e. if you have
\usepackage{jinstpub} % for details on the use of the package, please
\usepackage{indentfirst}

\title{A fast spatial resolution optimizing method for track-ion using a GEM detector based on the time information}

\author[a]{Huiyin Wu,}
\author[a]{Jianjin Zhou,}
\author[b]{Herun Yang,}
\author[a,c]{Jianrong Zhou}
\author[b]{Chenggui Lu,}
\author[c]{Xin Zhao,}
\author[a,1]{Bitao Hu,\note{Corresponding author.}}
\author[a,2]{Yi Zhang,\note{Corresponding author.}}
\affiliation[a]{School of Nuclear Science and Technology, Lanzhou University, Lanzhou 730000, China}
\affiliation[b]{Institute of Modern Physics, Chinese Academy of Sciences, Lanzhou, 730000, China}
\affiliation[c]{Institute of High Energy Physics, Chinese Academy of Sciences, Beijing 100049, China}

\emailAdd{yizhang@lzu.edu.cn}

\abstract{In this work, a fast method of vertex reconstructing for incident ions in GEM detectors was proposed. As inspired by the Time Projection Chamber (TPC), the time information of the consecutive signal samples from Front End Electronics (FEE) was employed. To demonstrate the method, an experiment involving a 2D-readout THGEM detector with an APV-25 FEE was taken. With one experiment dependent parameter in the analysis, the proposed method led to a spatial resolution of 0.45 mm, compared with the same number of 9.10 mm from the traditional center of gravity method.}


\arxivnumber{1234.56789} % only if you have one

\proceeding{N$^{\text{th}}$ Workshop on X\\
  when\\
  where}

\begin{document}
\maketitle
\flushbottom

\section{Introduction}
GEM (gas electron multiplier) detectors are widely used in high-energy physics and nuclear physics as a spatial sensitive detector to detect both photons and charged particles due to its good performance (high count rate, high effective gain, good spatial resolution, good plasticity)~\cite{a,b,c}. For photons, its spatial resolution can easily reach below 100 um with the traditional center of gravity method~\cite{d}. For charged particles, the spatial resolution is much worse, because of the different way of energy deposition. In the case of charged particles, charged particles produce tracks with ion-electron pairs, and the center of gravity of the track cannot correctly represent the incident vertex of the ion, especially when ion entries into the detector with a large angle. A TPC-like analysis method could be used to reconstruct the incident vertex~\cite{e,f}.
\par
In recent years, some fast on-line reconstruction technologies based on FPGA chip have been carried out. Limited by the ability of FPGA to handle non-logical operations, it is difficult to achieve complex reconstruction algorithm~\cite{g}. Therefore, the reconstruction algorithm is often simplified to reduce the consumption of precious chip resources. In this paper, we proposed a fast method of ion incident vertex reconstruction, only one parameter depends on experimental conditions is used to reconstruct the incident vertex of ions under the same experimental conditions.
\section{Reconstruction Principle}
\begin{figure}[htbp]
\centering % \begin{center}/\end{center} takes some additional vertical space
\includegraphics[width=.6\textwidth]{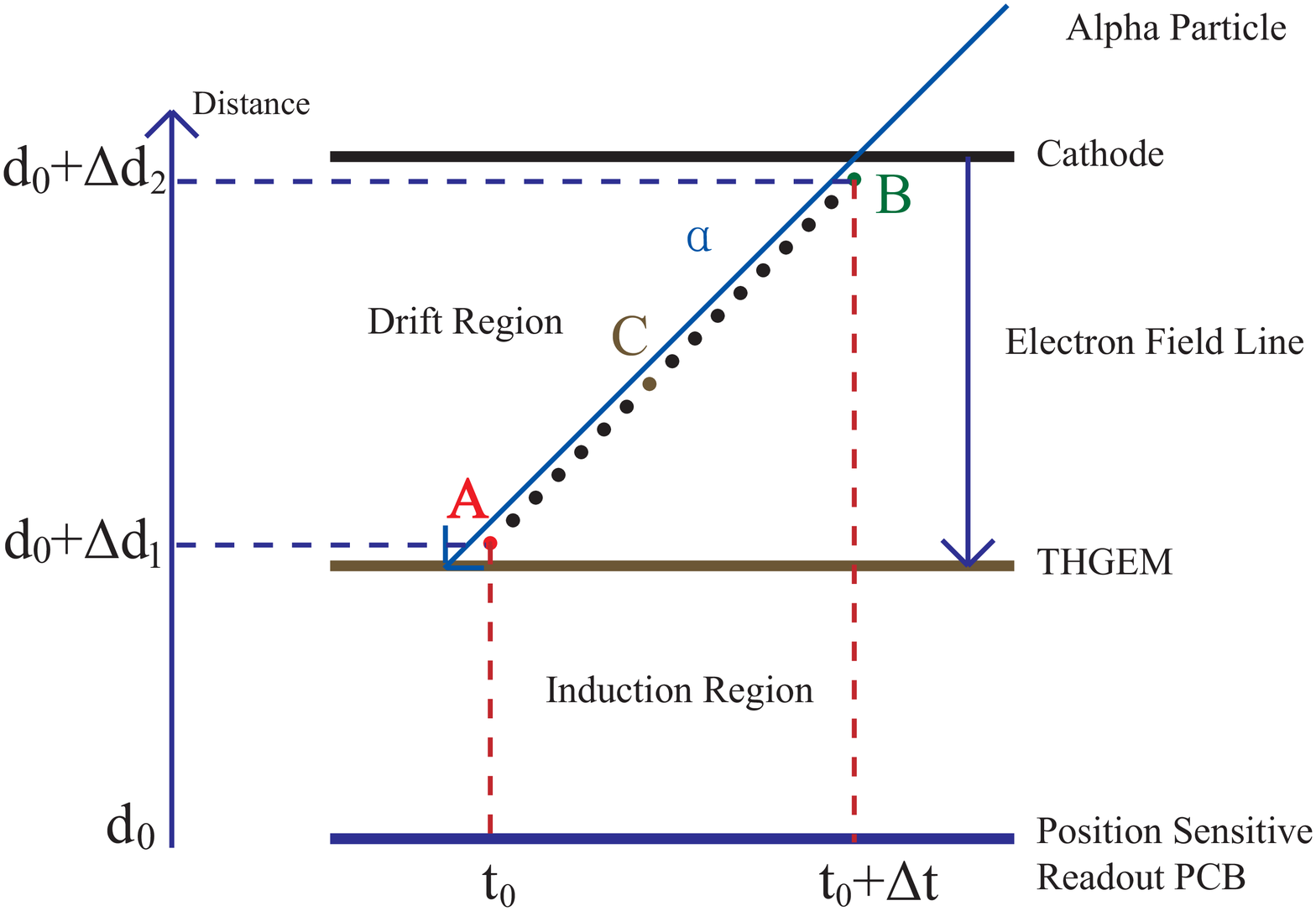}
\caption{\label{fig:1} Reconstruction principle. Point A is the end of the track and fires the readout PCB firstly, while point B which fires the readout PCB finally lastly is the beginning of the track. The relative time between point A and point B can be used to determine the incident vertex with the sampling integrated pre-amplifier.}
\end{figure}

The principle of vertex reconstruction is illustrated by the example of an alpha ion showed in Figure~\ref{fig:1}.The alpha ion entries with an angle and generates a track with ion-electron pairs in the drift region of the GEM detector. The velocity of the alpha ions with few MeV energy is much larger than the drift velocity of the ionized electrons in the detector, so with respect to the drift of electrons, the track can be considered to be produced instantly. The electrons at different positions of the track drift along the direction of the electric field line with slight lateral diffusion. The electrons at the end of the track (point A in Figure~\ref{fig:1}) multiplied by the GEM foil entry into the induction region and fire the strips firstly. On the contrary, the electrons at the beginning of the track (point B in Figure~\ref{fig:1}) fly through the drift region and entry into the induction region and fire the strips finally. The time information of the track can be used to achieve the ion incident vertex reconstruction. The typical signal duration time is 300 ns. APV-25 front-end-electronics with 25 ns sampling period samples the signal on each strip~\cite{h,i}. 30 consecutive samples can be acquired by the data acquisition system with a typical rise time of 150 ns, which allow to record the whole waveform of the alpha ion.

 \section{Experimental Setup}
\begin{figure}[htbp]
\centering % \begin{center}/\end{center} takes some additional vertical space
\includegraphics[width=.6\textwidth]{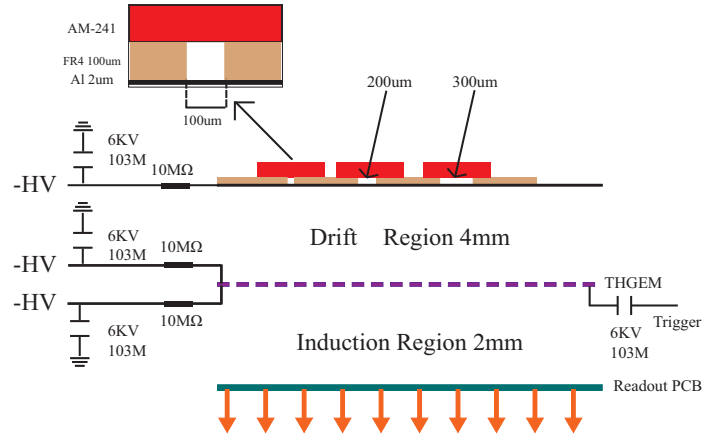}
\caption{\label{fig:2} Experimental setup. Two-dimensional position sensitive THGEM detector with one THGEM foil and 600 $\mu$m pitch in each dimensional. Alpha ions from the $^{241}$Am source entry drift region with a maximum angle as $71^{\circ}$. The signal on the lower side of THGEM foil provided a trigger signal for the DAQ system.}
\end{figure}

To illustrate the reconstruction principle, A single-layer THGEM detector was developed, which worked in a proportional mode as shown in Figure~\ref{fig:2}. It consisted of one THGEM foil (Au-coated with a thickness of 200 um, pitch of 600 um, a hole diameter of 200 um, and a rim of 80 um), a cathode plane and a read-out anode with the pitch of 600 um in each dimensional~\cite{j}. The intrinsic spatial resolution of the detector was about 173 um (600/$\sqrt{12}$ um). The detector operated with a continuously flushed Ar/CO$_{2}$ gas mixture (90/10 percentage in volume) and had a square sensitive area of 100 mm $\times$100 mm, a 4-mm drift region, and a 2-mm induction region. Three voltage dividers were employed to supply an electric field of to the detector, three resistances and capacitances were used to reduce the noise due to the voltage dividers. The work bias voltages were -1830V, -1350 V and -600 V, respectively. The voltage on the THGEM was 750 V, the electric field of drift region and induction region were 1200 V/cm and 3000 V/cm, respectively.
\par
The cathode was made of a 2-$\mu$m thick aluminized Mylar foil, which minimized the energy loss of alpha ions from a $^{241}$Am source to make the alpha ions run through the entire drift region. A 100-um thick FR4 foil, which could completely block the alpha ions except three slits on it, was placed on the cathode. The widths of the slits were 100 $\mu$m, 200 $\mu$m, and 300 $\mu$m, respectively. The distance between the slits was 30 mm. Three alpha sources were placed right on the slits, and an alpha ion could entry drift region at a maximum of $71^{\circ}$. The readout system was built on APV-25 front-end boards and a FPGA-based data acquisition system reading 30 consecutive samples in one trigger signal. The velocity of a 5.29 MeV alpha ion was about 15.9 mm/ns given by LISE++~\cite{k}, and the velocity of electrons in the drift region was about 0.05 mm/ns which was calculated by Garfield++~\cite{l}. Compared with the electron drifting time, the time of generating a track was negligible. Thus, it was reasonable to assume that all electrons in a track were generated and started to drift at the same time.
\section{Results and discussion}
\begin{figure}[htbp]
\centering % \begin{center}/\end{center} takes some additional vertical space
\includegraphics[width=.8\textwidth]{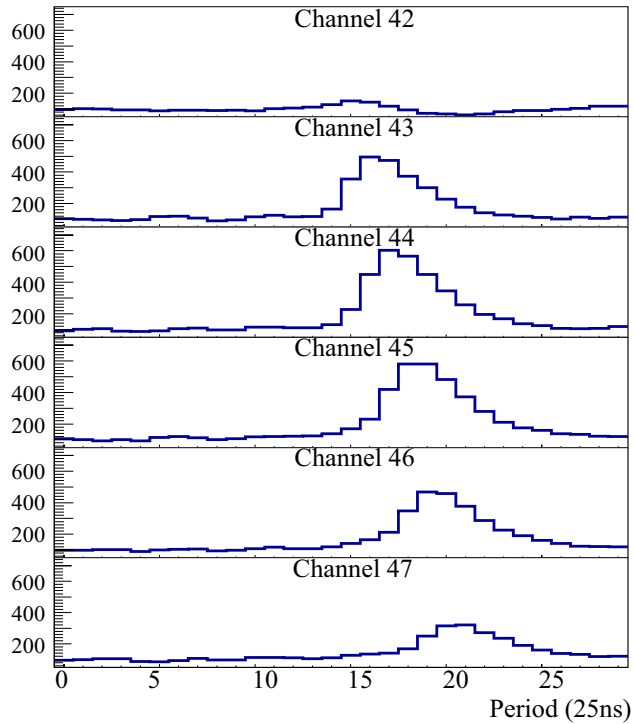}
\caption{\label{fig:3}  The signals on the 6 adjacent electrodes, it is obvious that the signals on different electrodes come in different time. }
\end{figure}

For an alpha track, it induces signals in several adjacent electrodes, in which the signals are sampled by the FEE in a frequency of 40 MHz, as shown in Figure~\ref{fig:3}. In Figure~\ref{fig:3} different histograms show the signals on the 6 adjacent electrodes. For one electrode, the signal is sampled for 30 FEE working periods. It is obvious that the signals on different electrodes come in different time. The electrode 43 gets the signal earliest, whose signal corresponds to the electrons ionized at the end of the track. Accordingly, the electrode 48 gets the signal latest, whose signal corresponds to the electrons ionized at the beginning of the track.
\par
\begin{figure}[htbp]
\centering % \begin{center}/\end{center} takes some additional vertical space
\includegraphics[width=.6\textwidth]{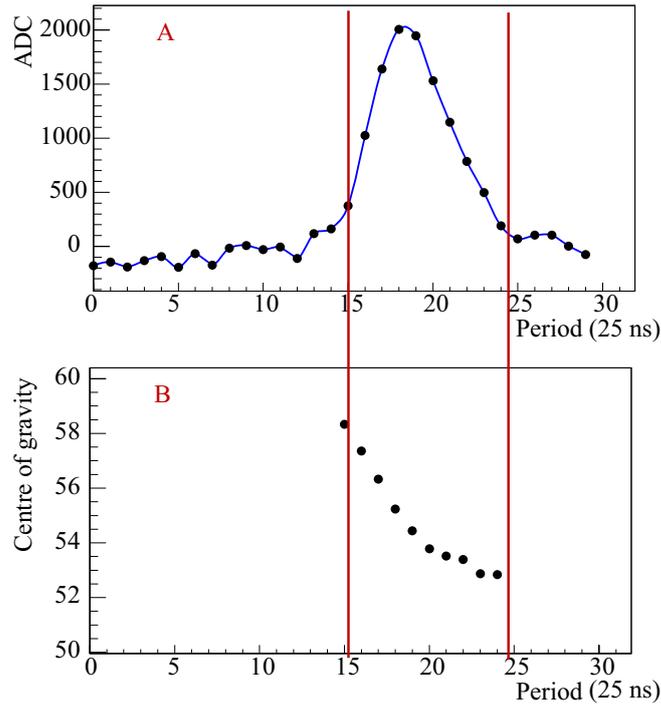}
 \caption{\label{fig:4} A typical signal of an alpha ion. A: Signal waveform of an alpha ion. B: Samples order against the gravity center of each sample. The center of gravity gradually approaches the incident vertex (48.5). But there still is a deviation.}
\end{figure}

Since the entire signal induced by a proton track lasts for several FEE periods, within each FEE period it is only a part of the signal which DAQ deals with. To correctly reconstruct the vertex of a track, it is necessary to firstly determine which FEE periods contains the entire signal of a track. Figure~\ref{fig:4} shows the signals of a track sampled in several FEE periods. The upper panel shows the sum of signal amplitude of all electrodes within one FEE period. As a constant background subtraction was applied, the sum of signal amplitude in the FEE periods in which there is no real signal but noises varies around 0. In Figure~\ref{fig:4}A it is clear that the signal starts in the period 15 and ends around the period 24, as marked by the two red lines. In the experiment, we put an amplitude cut on each FEE period to determine if there are signals induced by a track. The lower panel shows the centers of gravity of the signals in each FEE period. It is clearly shown that the centers of gravity shifts among FEE periods, which is consistent with Figure~\ref{fig:3}. In the experiment according to the position of the slit from which the track shown in Figure~\ref{fig:4} flies in, the vertex of the track should located around 48.5 on the y-axis in Figure~\ref{fig:4}B. In Figure~\ref{fig:4}B the centers of gravity of signal shifts toward to the vertex. It is an illustration of the reconstruction principle in Section 2. However, due to the finite time response of the FEE and DAQ, the signal center of gravity in the last FEE period cannot accurately represent the incident vertex. This effect is also observed in Figure~\ref{fig:4}B. At the end of the track, the center of gravity is the end of the track, because the other channels are not fired yet. But at the beginning of the track, the channels near the end of the track have been fired and influence calculating the center of gravity, which causes the center of gravity to shift towards the end of the track. The center of gravity of each sample can not accurately represent the incident vertex.
\par
\begin{figure}[htbp]
\centering % \begin{center}/\end{center} takes some additional vertical space
\includegraphics[width=.8\textwidth]{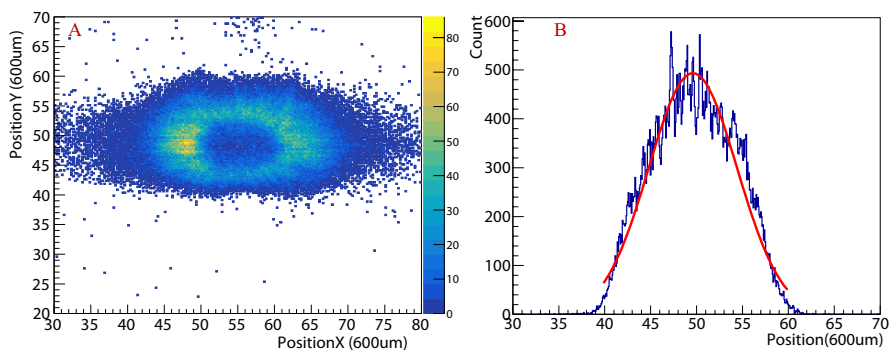}
\caption{\label{fig:5} A: Position spectrum in two-dimensional. B: Position spectrum in x dimensional.}
\end{figure}

In a normal analysis, the track vertex is determined by the center of gravity of all the signals induced by the entire track. For the track shown in Figure~\ref{fig:4}B this determined result should be a position around 55, which would have an obvious deviation with the real vertex. As a comparison, the result of this analysis method on the experimental data is shown in Figure~\ref{fig:5}. Although in the experiment the alpha particles can fly into the detector through 3 slits, only the data from one slit are plotted in Figure~\ref{fig:5}. In Figure~\ref{fig:5} the left panel shows the 2-D distribution of the track vertexes determined by the center of gravity. As all the alpha particles in the figure are from one slit, the real vertex distribution should represent the shape of the slit. The obvious difference between the pattern shown in Figure~\ref{fig:5}A and the shape of the slits clearly shows the failure of the normal analysis method. Figure~\ref{fig:5}B shows the center of gravity distribution in x direction which is perpendicular to the direction of the slits. In order to evaluate the position resolution of this normal analysis method, a convolution of a rectangular function with a Gaussian function is applied as a fitting function on the histogram shown in Figure~\ref{fig:5}B. The fitting result is shown as the red curve in Figure~\ref{fig:5}B and the extracted position resolution is 9.1 mm.

\par
\begin{figure}[htbp]
\centering % \begin{center}/\end{center} takes some additional vertical space
\includegraphics[width=.6\textwidth]{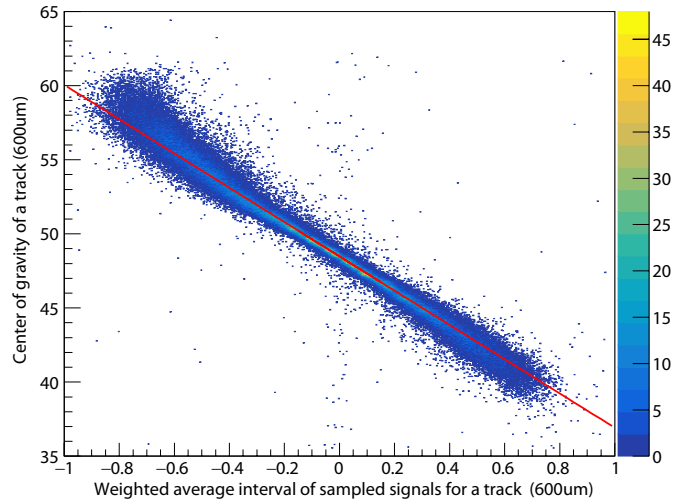}
\caption{\label{fig:6} Weighted average interval of samples versus center of gravity for a track. There is a obvious linear correlation.}
\end{figure}

As illustrated in Figure~\ref{fig:4} and Figure~\ref{fig:5}, neither the center of gravity of an entire track nor the signal in any FEE period can not represent the track vertex, it is necessary to find another way to reconstruct the vertex. Figure~\ref{fig:6} shows a clear correlation between the weighted average center of the track for a slit and the weighted average center of gravity shift in one dimensional of track. The weighted average center of gravity shift means the weighted average of the center of gravity shift for two consecutive sampling periods and signal size, which is defined as:
\begin{equation}
\label{eq:y:3}
\frac{ \sum_{sa=i}^n{(\alpha_{sa+1}-\alpha_{sa}) \times (\beta_{sa+1}+\beta_{sa})}}{2 \times \sum_{sa=i}^n\beta_{sa} },
\end{equation}
where $\alpha_{sa}$ represents the center of gravity for the sample period \emph{sa}, and $\beta_{sa}$ represents the total signal for the sample period \emph{sa}.
\par
The ratio of the center of gravity to the weighted average center of gravity shift is a constant, which can be obtained from the linear fitting in Figure~\ref{fig:6}. Here, the value is -11.55, that means for each track, its reconstructed vertex is determined by the following formula:
\begin{equation}
\label{eq:y:3}
 Position =\emph{A} + 11.55 \times \emph{B},
\end{equation}
where \emph{A} represents the weighted average center of track, and \emph{B} represents the weighted average center of gravity shift.

\par
\begin{figure}[htbp]
\centering % \begin{center}/\end{center} takes some additional vertical space
\includegraphics[width=.8\textwidth]{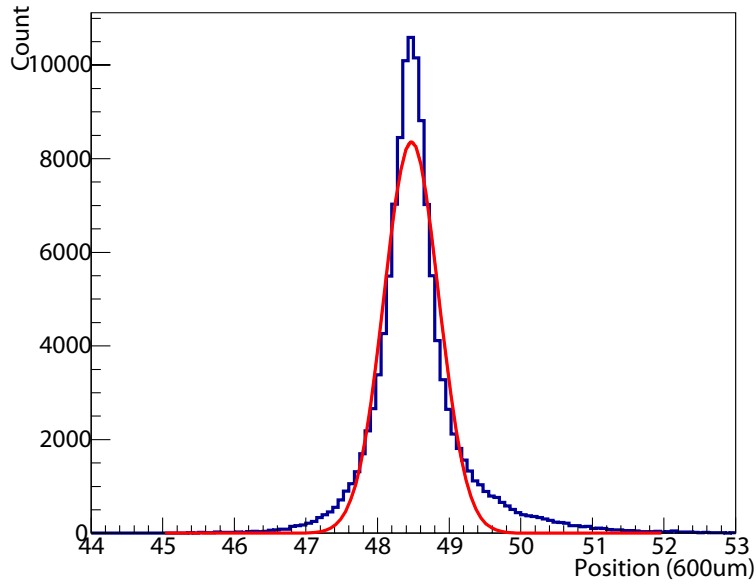}
\caption{\label{fig:7} Spatial resolution after corrected,the resolution is a FWHM of 0.45 mm achievable by the convolution with the Gaussian-distribution method }
\end{figure}

 The spatial resolution after correction with Eq.(4.2) is shown in Figure~\ref{fig:7}, the resolution is a FWHM of 0.45 mm achievable by the convolution with the Gaussian-distribution method. Being compared to the traditional method, this method has an order-of-magnitude optimization in spatial resolution.

\par
\begin{figure}[htbp]
\centering % \begin{center}/\end{center} takes some additional vertical space
\includegraphics[width=.8\textwidth]{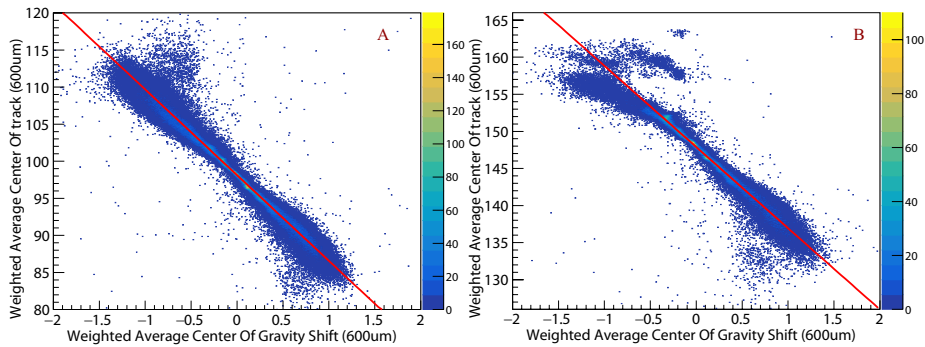}
\caption{\label{fig:8} A: Weighted average center of gravity shift against weighted average center of track for the middle silt.  B: Weighted average center of gravity shift against weighted average center of track for the right silt}
\end{figure}

The same processing is performed on the ions emitted from the other two slits. The results are shown in Figure~\ref{fig:8}. The parameters obtained by fitting the data of the two slits are 11.43 (Figure~\ref{fig:8}A) and 11.40 (Figure~\ref{fig:8}B), respectively. Distortion signals are removed when processing the right slit data due to the distortion of the detector's drift field in fringe region. The parameters of the three slits are almost the same under the same working condition, indicating that for any track under this experimental condition, such a phenomenological parameter can be determined and applied to reconstruct the incident vertex. The main factor affecting this parameter is the electron drift velocity, which depends on the electric field of the drift region. If the electron drift velocity becomes faster, the electronic response is relatively slow, and the center of gravity movement of each sampling period becomes smaller, which leads to a larger parameter. This parameter is independent to the incident angle of the tracks, the counting rate, the slit width, etc.
\par
\begin{figure}[htbp]
\centering % \begin{center}/\end{center} takes some additional vertical space
\includegraphics[width=.8\textwidth]{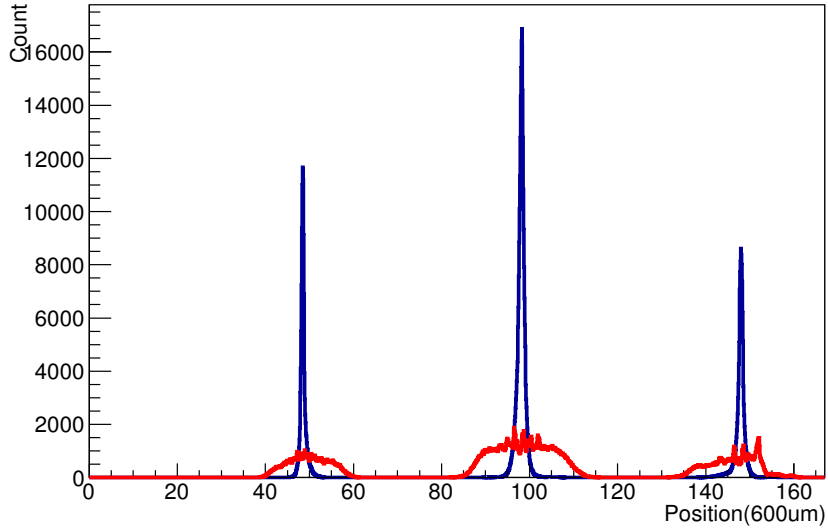}
\caption{\label{fig:9} Alpha ion reconstructed position spectrum of three slits using the center of gravity technique (in red) and the TPC technique (in blue). }
\end{figure}

The correction parameter of the first slit is used to reconstruct the incident vertex of tracks from all three slits. The result is shown in Figure~\ref{fig:9}. The center of three slits are located at 48.47 (600 um), 98.10 (600 um) and 148.02 (600 um), respectively. The distances between the slits are 29.78 mm and 29.98 mm. Compared with the physical value of the distances, which is 30 mm, this impressive coincidence shows an unambiguous evidence of the proposed reconstruction method.

\section{Conclusion}
In this work, a fast method of vertex reconstructing for incidental ions in GEM detectors is proposed. The analysis shows that with an experiment dependent parameter, the method offers a spatial resolution of 0.45 mm. Compared with a result offered by the traditional gravity method, which is 9.10 mm, there is an improvement over one order of magnitude. The parameter, which depends on the configuration of the detector and is independent of the incident track, can be extracted by a linear fit of predefined signals as a calibration. Experimental evidence that unambiguously proves the capability of this method is described.
\par
As discussed before, the analysis in this work assumes the incidental tracks generate signals in a linear manner. Higher order effects are ignored. It is possible to achieve a higher precision by taking the un-linear effects into account. But it will make the reconstruction much more complicated. As also discussed before, current method eliminates the complexity so that it is possible to be implemented in a FPGA-based on-line DAQ system. It is another considerable advantage of the reconstruction method proposed in this work.

\acknowledgments
This work was supported by the National Natural Science Foundation of China (Grant Nos.11405077, No.11575073, No.11575268 and No.11405233) and the Fundamental Research Funds for the Central Universities (Grant No.lzujbky-2018-71).


\end{document}